\documentclass[aps,prl,floatfix,showpacs,twocolumn]{revtex4-1}
\usepackage{ulem}
\usepackage{dcolumn}
\usepackage{bm}
\usepackage{color}
\usepackage{amssymb}
\usepackage{amsmath}
\usepackage{graphicx}
\usepackage{amsfonts}
\usepackage{slashed}
\usepackage{pstricks}
\usepackage{float}
\usepackage{hyperref}
\usepackage{array}
\allowdisplaybreaks
\usepackage{dcolumn}
\usepackage{epsf}
\begin{document}
\title{Electromagnetic response of $^{12}$C: a first-principles calculation}
\author{
A.\ Lovato$^{\, {\rm a} }$,
S.\ Gandolfi$^{\, {\rm b} }$,
J.\ Carlson$^{\, {\rm b} }$,
Steven\ C.\ Pieper$^{\, {\rm a} }$,
and R.\ Schiavilla$^{\, {\rm c,d} }$
}
\affiliation{
$^{\,{\rm a}}$\mbox{Physics Division, Argonne National Laboratory, Argonne, IL 60439}\\
$^{\,{\rm b}}$\mbox{Theoretical Division, Los Alamos National Laboratory, Los Alamos, NM 87545}\\
$^{\,{\rm c}}$\mbox{Theory Center, Jefferson Lab, Newport News, VA 23606}\\
$^{\,{\rm d}}$\mbox{Department of Physics, Old Dominion University, Norfolk, VA 23529}
}
\date{\today}

\begin{abstract}
The longitudinal and transverse electromagnetic response functions of
$^{12}$C are computed in a ``first-principles'' Green's function Monte Carlo
calculation, based on realistic two- and three-nucleon interactions and
associated one- and two-body currents.  We find excellent agreement
between theory and experiment and, in particular, no evidence for the
quenching of measured versus calculated longitudinal response.
This is further corroborated by a re-analysis of the Coulomb sum rule,
in which the contributions from the low-lying $J^\pi\,$=$\, 2^+$, $0^+_2$ (Hoyle),
and $4^+$ states in $^{12}$C are accounted for explicitly in evaluating
the total inelastic strength.
\end{abstract} 


\index{}\maketitle
One of the challenges in quantum many-body physics is calculating the electroweak
response of a nucleus by fully accounting for the dynamics of its constituent
nucleons.  In this paper we report the first such calculation for the
electromagnetic response of the $^{12}$C nucleus.  

The nucleons interact with each other via two- and three-body forces and
with external electroweak fields via one- and two-body,
and smaller many-body, currents.  This dynamical picture
of the nucleus in which the consequences of the
nucleons' substructure on its structure and response are subsumed into effective many-body
forces and currents is by now well established.  When coupled to numerically exact
methods, such as the Green's function Monte Carlo (GFMC) methods adopted in this work, it
has led to a quantitative and successful ``first-principles'' understanding of many nuclear
properties: the low-lying energy spectra of nuclei up to $^{12}$C~\cite{Carlson:2015}
(and references therein); their radii and magnetic moments~\cite{Pastore:2013,Pastore:2014};
their elastic and inelastic electromagnetic
form factors~\cite{Wiringa:1998,Lovato:2013}; electroweak transitions between their low-lying states
($M1$ and $E2$ widths~\cite{Pastore:2013,Pastore:2014}, and $\beta$-decay and
electron-capture rates~\cite{Schiavilla:2002}); properties of their
ground-state structure, such as the momentum distributions of nucleons and nucleon pairs~\cite{Wiringa:2014};
insights into the role that the dominant features of the nuclear interaction--the short-range
repulsion and long-range tensor nature--have in shaping their ground-state structure~\cite{Schiavilla:2007};
and more (for a recent review see~\cite{Carlson:2015}).
One of the key features of this approach is the assumption that the couplings
of the external fields to the nucleons are governed by those in
free-space with modifications induced primarily by two-nucleon currents.

Here we report calculations of the $^{12}$C electromagnetic longitudinal and transverse response
functions, denoted respectively as $R_L(q,\omega)$ and $R_T(q,\omega)$, where
$q$ and $\omega$ are the electron momentum and energy transfers.  These response
functions are obtained experimentally by Rosenbluth separation of inclusive $(e,e^\prime)$
scattering data~\cite{Barreau:1983s,Jourdan:1996}.  The calculations are based on the AV18+IL7 combination
of two and three-nucleon potentials~\cite{Wiringa:1995,Pieper:2008}
and accompanying set of two-body electromagnetic currents
(for a review see~\cite{Carlson:2015} and references therein).   GFMC methods are used to
compute these responses as functions of imaginary time~\cite{Carlson:1992,Carlson:1994},
and maximum-entropy techniques to infer from these imaginary-time data the actual $R_L(q,\omega)$ and
$R_T(q,\omega)$~\cite{Bryan:1990,Jarrell:1996,Lovato:2015}.  These latter two aspects
of this study are discussed below.

Accurate calculations of the nuclear response are necessary to reliably
test this realistic framework of nuclear dynamics.
In simplified approaches, for example,
an increase in nucleon size has been advocated to explain the
depletion of the nuclear structure functions measured by deep inelastic scattering (the EMC
effect~\cite{Close:1983}), 
the quenching of the quasi-elastic longitudinal response measured in
$(e,e^\prime)$ scattering off light and heavy nuclear targets~\cite{Noble:1981,Cloet:2016},
and the suppression in the ratio of transverse
to longitudinal polarization transfers in $^4$He relative to the ratio in hydrogen, measured
via the $^4$He$(\vec{e},e^\prime \vec{p}\,)^3$H reaction at quasi-elastic kinematics at
Jefferson Lab~\cite{Strauch:2003s} (and references therein).

Clearly the question of in-medium modifications is model dependent.  Indeed,
theoretical approaches based on the realistic picture outlined above
indicate that binding and correlation effects, included by employing realistic
spectral functions, lead to average removal energies much larger than
those adopted in standard EMC calculations, and provide a quantitative account
of both the size and density dependence of the EMC effect~\cite{Benhar:1997,Benhar:1999,Benhar:2012b}. 
Such approaches also show that spin-dependent final state interaction effects and corrections
beyond the impulse approximation, induced by two-body electromagnetic
currents, resolve the discrepancy between theory and experiment in the case
of the polarization-transfer ratio when the free nucleon electromagnetic form
factors are used in the nuclear currents~\cite{Schiavilla:2005}.

The quark-meson coupling approach, which attempts to self-consistently account for nucleon and nuclear
structure~\cite{Lu:1999,Guichon:2004}, leads to a reduction of the proton electric form factor, and, as a
consequence, to a significant quenching of  the longitudinal response function of nuclear
matter and associated Coulomb sum rule~\cite{Cloet:2016}.  Such a model does
not explain the large enhancement of the transverse response or the momentum-transfer dependence
in the quenching of the longitudinal one.  It should also be noted that medium modifications
are not an inevitable consequence of the quark substructure of the nucleon.  For example,
a study of the two-nucleon problem in a flux-tube model of six quarks interacting via single
gluon and pion exchanges~\cite{Paris:2000} indicates that the nucleons retain their individual
identities down to very short separations, with little distortion of their substructures.

In this paper we show that accurate calculations of the response based on a
realistic correlated nuclear wave function and containing one- and two-body
currents with free nucleon form factors can completely reproduce the
$^{12}$C longitudinal and transverse electromagnetic response below the
delta resonance.

The longitudinal and transverse response functions are defined as
\begin{align}
R_\alpha (q,\omega)&=\sum_f \langle f | j_\alpha({\bf q},\omega) |0\rangle 
\langle f | j_\alpha({\bf q},\omega) |0\rangle^* \nonumber\\
& \times \delta(E_f-\omega-E_0)\ ,\qquad \alpha=L,T
\label{eq:res_def}
\end{align}
where $|0\rangle$ and $|f\rangle$ represent the nuclear initial and final
states of energies $E_0$ and $E_f$, and $j_L({\bf q},\omega)$ and $j_T({\bf q},\omega)$ are the
electromagnetic charge and current operators, respectively.
A direct calculation of $R_\alpha(q,\omega)$ is impractical, because it would require
evaluating each individual transition amplitude $|0\rangle \longrightarrow |f\rangle$
induced by the charge and current operators.  To circumvent this
difficulty, the use of integral transform techniques has proved to be quite helpful.
One such approach is based on the Laplace transform of $R_\alpha(q,\omega)$---i.e.
the Euclidean response~\cite{Carlson:1992} defined as
\begin{equation}
E_\alpha(q,\tau)=\int_{\omega_{\rm el}^+}^\infty d\omega\, e^{-\omega\tau} \frac{R_\alpha(q,\omega)}
{[G_E^p(q,\omega)]^2} \ ,
\label{eq:euc_def}
\end{equation}
where $G_E^p(q,\omega)$ is the (free) proton electric form factor and the integration
excludes the contribution due to elastic scattering ($\omega_{\rm el}$ is the energy
of the recoiling ground state).  We elaborate this issue further below;
for now it suffices to note that, in the specific case of $^{12}$C, the ground state has
quantum numbers $J^\pi\,$=$\,0^+$ and therefore the elastic contribution vanishes
in the transverse channel. With the definition given in Eq.~(\ref{eq:euc_def}),
 the Euclidean response function above can be thought of as being due to point-like,
but strongly interacting, nucleons, and can simply be expressed as
\begin{equation}
E_\alpha(q,\tau)\!=\!\langle 0| O^\dagger_{\alpha}({\bf q}) e^{-(H-E_0)\tau} 
O_\alpha({\bf q}) |0\rangle
-|F_\alpha(q)|^2 {\rm e}^{-\tau \omega_{\rm el}} \ ,
\label{eq:euc_me}
\end{equation}
where $H$ is the nuclear Hamiltonian (here, the AV18+IL7 model),
$F_\alpha(q) =\langle0| O_\alpha({\bf q})|0\rangle$
is the elastic form factor, and
in the electromagnetic operators $O_\alpha({\bf q})$ the dependence on the
energy transfer $\omega$ has been removed by dividing the current
$j_\alpha({\bf q}, \omega)$ by $G_E^p(q,\omega)$~\cite{Lovato:2015}.
The calculation of this matrix element is then carried out with
GFMC methods~\cite{Carlson:1992} similar to those used in projecting out
the exact ground state of $H$ from a trial state~\cite{Carlson:1987}.
It proceeds in two steps.  First, an unconstrained imaginary-time
propagation of the state $|0\rangle$ is performed
and saved.  Next, the states $O_{\alpha}({\bf q}) |0\rangle$ are evolved in imaginary
time following the path previously saved.  During this latter imaginary-time evolution, scalar
products of ${\rm exp}\left[-\!\left(H\!-\!E_0\right) \tau_i\right]O_{\alpha}({\bf q})| 0\rangle$
with $O_{\alpha}({\bf q})|0\rangle$ are evaluated on a grid of $\tau_i$ values, and
from these scalar products estimates for $E_{\alpha}(q,\tau_i)$ are
obtained (a complete discussion of the methods is in
Refs.~\cite{Carlson:1992,Carlson:2002}).
We use our best variational trial wave function $\Psi_T$ for $|0\rangle$ and thus the response
functions are those of $\Psi_T$ instead of the evolved GFMC wave function.
The sum rule results of  Ref.~\cite{Lovato:2013} suggest that this a
good approximation.

Following Ref.~\cite{Lovato:2015} (see also extended material submitted
in support of that publication), we have exploited maximum entropy
techniques~\cite{Bryan:1990,Jarrell:1996} to perform the analytic continuation
of the Euclidean response function---corresponding to the {\it inversion} of the 
Laplace transform of Eq. (\ref{eq:euc_def}).   However, we have improved
on the inversion procedure described in~\cite{Lovato:2015} in order to better
propagate the statistical errors associated with $E_\alpha(q,\tau)$
into $R_\alpha(q,\omega)$. Specifically, the smallest possible
value for parameter $\alpha$ (see Ref.~\cite{Lovato:2015}) has been chosen to perform a first inversion
of the Laplace transform, which is then independent of the prior. The resulting
response function $R^{(0)}$ is the one whose Laplace transform $E^{(0)}$ is
the closest to the original average GFMC Euclidean response. Then, 100
Euclidean response functions are sampled from a multivariate Gaussian
distribution, with mean value $E^{(0)}$ and covariance estimated from the
original set of GFMC Euclidean responses. The corresponding response
functions, obtained using the so called ``historic maximum entropy'' technique~\cite{Jarrell:1996},
are used to estimate the mean value and the variance of the final inverted response function.

\begin{table}[bth]
\vspace{0.3cm}
{\renewcommand{\arraystretch}{1.0}
\renewcommand{\tabcolsep}{0.2cm}
\begin{tabular}{c | c c c c} 
 $q$ (MeV/c) 	& $2^+$ & $0^+_2$  & $4^+$\\
\hline
\rule{0pt}{2.5ex}
$300$ & 0.128 & 0.0313 & 0.0010\\ 
$380$ & 0.0743 & 0.0052 & 0.0012\\ 
$570$ & 0.0043 & 0.0045 & 0.00059 \\
\end{tabular} }
\vspace{0.1cm}
\caption{Measured longitudinal transition form factors, defined as $\langle f | O_L({\bf q})|0\rangle/Z$,
to the $f\,$=$\,2^+$, $0^+_2$ (Hoyle), and $4^+$ states in $^{12}$C.  Experimental data are from
Refs.~\cite{Nakada:1971a,Nakada:1971b,Chernykh:2010}, and have been divided by the
proton electric form factor $G_E^p(q,\omega_f)$ with $\omega_f=E_f-E_0$, as described in Ref. \cite{supplemental}. }
\label{tab:ff}
\end{table}
We now proceed to address the issue of excluding the elastic contribution.
The low-lying excitation spectrum of $^{12}$C consists of $J^\pi\,$=$\, 2^+$, $0^+_2$ (Hoyle), and
$4^+$ states with excitation energies $E^\star_f-E_0$ experimentally known to be, respectively,
4.44, 7.65, and 14.08 in MeV units~\cite{Ajzenberg:1990}.
The contributions of these states to the quasi-elastic longitudinal and transverse response
functions extracted from inclusive $(e,e^\prime)$ cross section measurements
are not included in the experimental results.  
Therefore, before comparing experiment with the present theory, which
computes the {\it total} inelastic response rather than just the {\it quasi-elastic} one,  
we need to remove these contributions explicitly.  This
is simply accomplished by first defining 
\begin{equation}
\overline{E}_\alpha(q,\tau) = E_\alpha(q,\tau)  - \sum_{f}  |\langle f | O_\alpha({\bf q})|0\rangle|^2
\, {\rm e}^{-(E_f - E_0)/\tau} , 
\label{eq:etcr}
\end{equation}
where in the sum only the states $f\,$=$\, 2^+$, $0^+_2$, and $4^+$ are included, and
then inverting $\overline{E}(q,\tau)$ (the energies $E_f$ differ from
$E_f^\star$, since the former include recoil kinetic energies).  We do not
attempt a GFMC calculation of the
excitation energies of these states or associated transition form factors---it
would require explicit calculations of these states or
propagating ${\rm exp}\left[-\!\left(H\!-\!E_0\right) \tau\right]O_{\alpha}({\bf q})| 0\rangle$
to computationally prohibitive large values of $\tau$.  Rather, we use  the
experimental energies and form factors, listed in Table~\ref{tab:ff}, to
obtain $\overline{E}_\alpha(q,\tau)$ from the GFMC-calculated $E_\alpha(q,\tau)$.
Because of the fast drop of these form factors with increasing momentum transfer,
the correction in Eq.~(\ref{eq:etcr}) for the longitudinal
channel ($\alpha\, $=$\, L$ ) is significant at $q=300$ MeV/c, but completely negligible at $q=570$ MeV/c.
In the case of the transverse channel ($\alpha\,$=$\, T$), possible contributions from
$E2$ and $E4$ transitions to the $2^+$ and $4^+$ states are too small \cite{Flanz:1978,
Karataglidis:1995} to have an impact on $\overline{E}_T(q,\tau)$.

\begin{center}
\begin{figure}[bth]
\includegraphics[width=8.5cm]{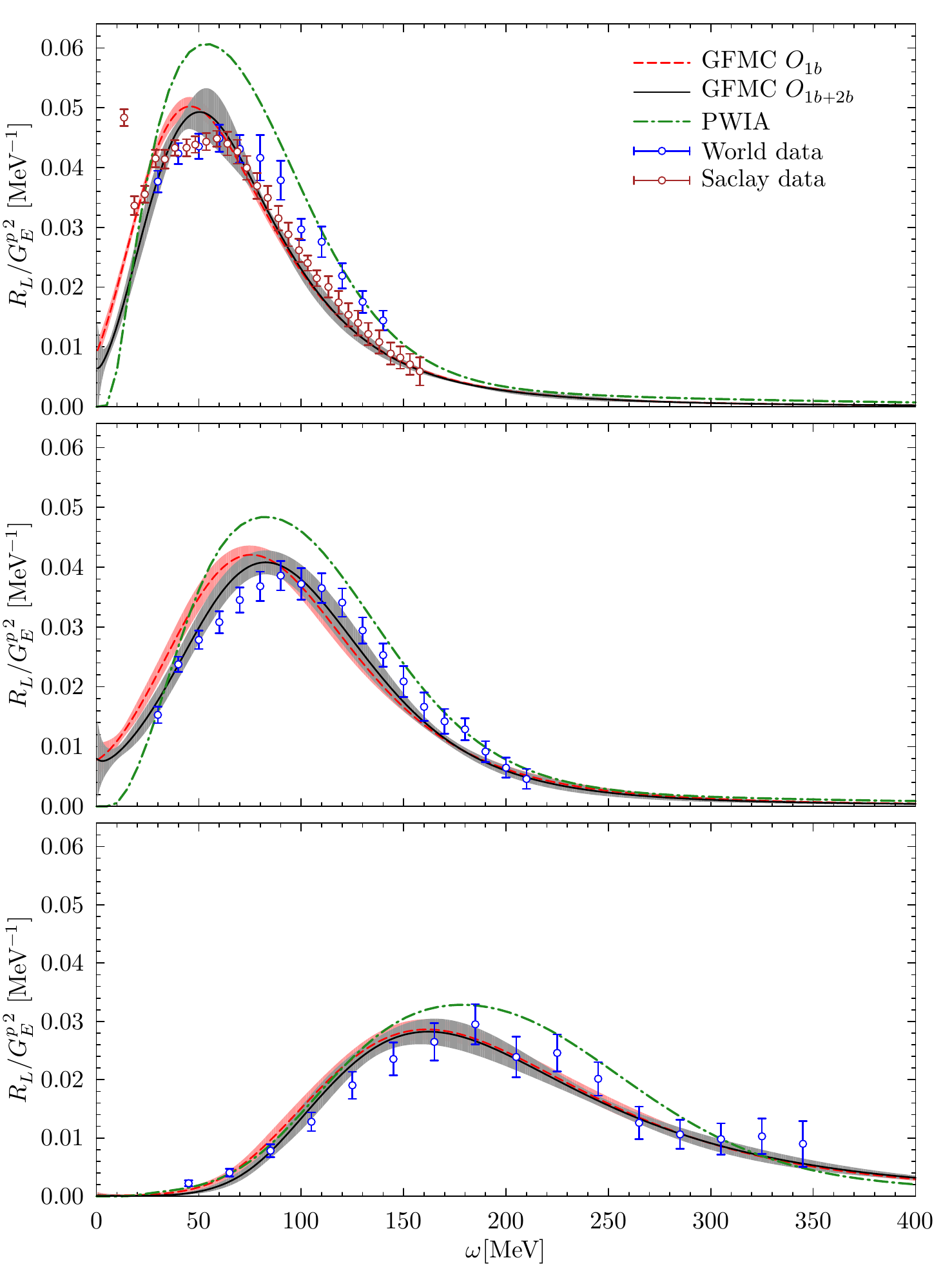}
\caption{(Color online) Electromagnetic longitudinal response functions of 
$^{12}$C for $q$ in the range (300--570) MeV.
Experimental data are from Refs. \cite{Barreau:1983s,Jourdan:1996}. See text for further explanations.}
\label{fig:f1}
\end{figure}
\end{center}

\begin{center}
\begin{figure}[bth]
\includegraphics[width=8.5cm]{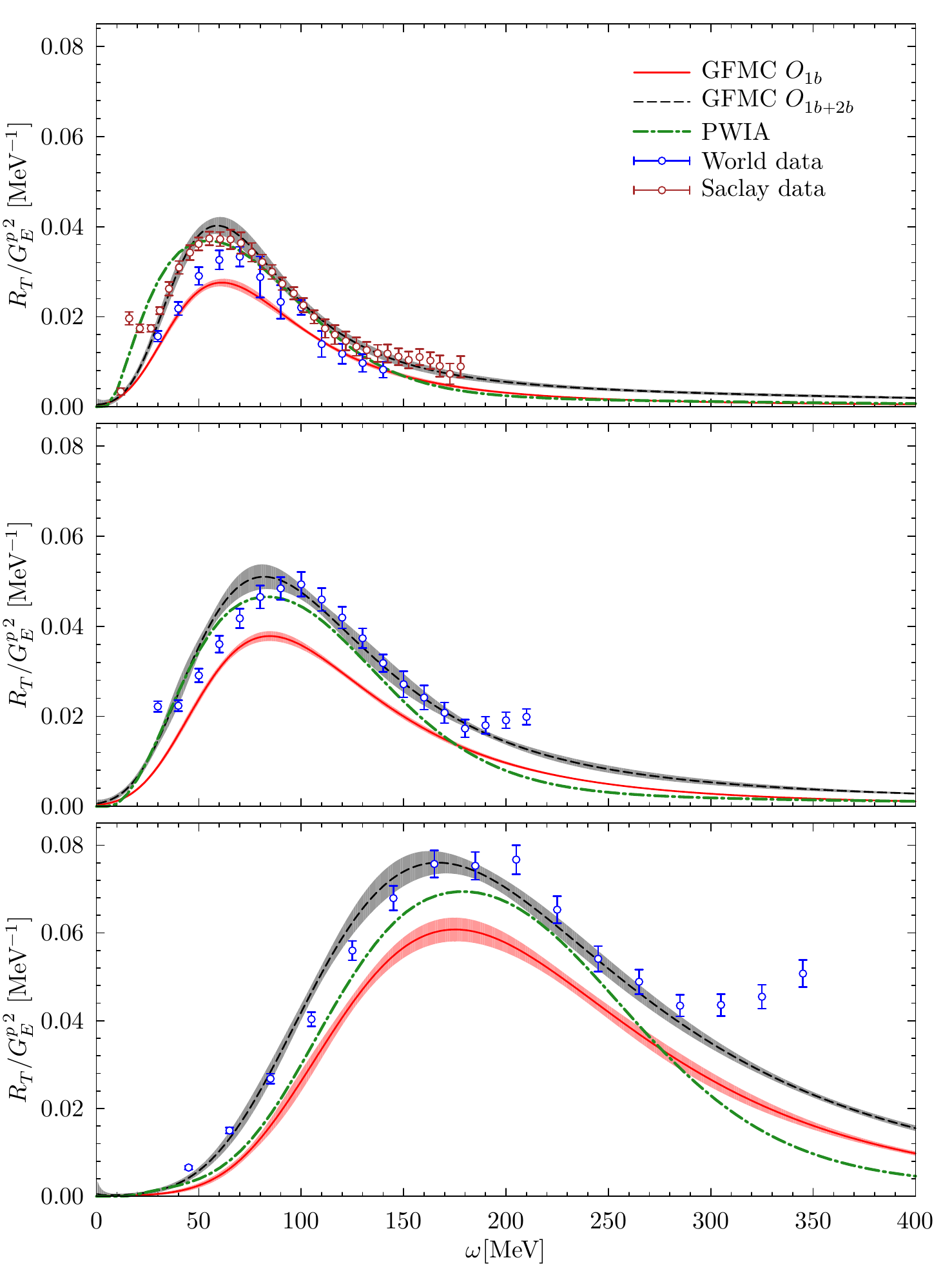}
\caption{(Color online) Same as Fig. \ref{fig:f1} but for the electromagnetic
transverse response functions.  Because pion production mechanisms
are not included, the present theory underestimates the (transverse)
strength in the $\Delta$ peak region, see in particular the $q\,$=$\, 570$ MeV/c case.}
\label{fig:f3}
\end{figure}
\end{center}

The longitudinal and transverse response functions obtained by maximum-entropy
inversion of the $\overline{E}_\alpha(q,\tau)$'s are displayed in Figs.~\ref{fig:f1} and~\ref{fig:f3},
respectively.  Theoretical predictions corresponding to GFMC calculations in which
only one-body terms or both one- and two-body terms are retained in the electromagnetic
operators $O_\alpha$---denoted by (red) dashed and (black) solid lines and labeled
GFMC-$O_{1b}$ and GFMC-$O_{1b+2b}$, respectively---are compared to the experimental response
functions determined from the world data analysis of Jourdan~\cite{Jourdan:1996} and,
for $q\,$=$\, 300$ MeV/c, from the Saclay data~\cite{Barreau:1983s}.  The (red and gray)
shaded areas show the uncertainty derived from the dependence of the 1b and 1b+2b results
on the default model adopted in the maximum-entropy inversion~\cite{Lovato:2015}.  This
uncertainty is quite small.  Lastly, the (green) dash-dotted lines correspond to 
plane-wave-impulse-approximation (PWIA)
calculations using the single-nucleon momentum distribution $N(p)$ of $^{12}$C obtained
in Ref.~\cite{Wiringa:2014} (see Ref.~\cite{Carlson:2015} for details on the PWIA calculation).

Figures~\ref{fig:f1}--\ref{fig:f3} immediately lead to the main conclusions of this
work: (i) the dynamical approach outlined above (with free nucleon electromagnetic
form factors) is in excellent agreement with experiment in both the longitudinal and
transverse channels; (ii) as illustrated by the difference between the
PWIA and GFMC one-body-current predictions
(curves labeled PWIA and GFMC-$O_{1b}$), correlations and interaction effects in the final
states redistribute strength from the quasi-elastic peak to the threshold and high-energy
transfer regions;  and (iii) while the contributions from two-body charge operators tend
to slightly reduce $R_L(q,\omega)$ in the threshold region, those from two-body currents
generate a large excess of strength in $R_T(q,\omega)$ over the whole $\omega$-spectrum
(curves labeled GFMC-$O_{1b}$ and GFMC-$O_{1b+2b}$), thus offsetting the quenching noted in (ii)
in the quasi-elastic peak.

As a result of this study, a consistent picture of the electromagnetic
response of nuclei emerges, which is at variance with the conventional
one of quasi-elastic scattering as being dominated by single-nucleon
knock-out.  This fact also has implications for the nuclear weak response probed
in inclusive neutrino scattering induced by charge-changing and neutral current
processes.  In particular, the energy dependence of the cross section
is quite important in extracting neutrino oscillation parameters.
An earlier study of the sum rules associated with the weak
transverse and vector-axial interference response functions in $^{12}$C
found~\cite{Lovato:2014} a large enhancement due to two-body currents in both the vector
{\it and} axial components of the neutral current.  Only neutral weak processes
have been considered so far, but one would expect these conclusions 
to remain valid in the case of charge-changing ones.  In this connection,
it is important to realize that neutrino and anti-neutrino cross sections differ
only in the sign of this vector-axial interference response, and that this
difference is crucial for inferring the charge-conjugation and parity violating
phase, one of the fundamental parameters of neutrino physics,
to be measured at the Deep Underground Neutrino Experiment (DUNE)\cite{lbne}.

\begin{center}
\begin{figure}[bth]
\includegraphics[width=8.5cm]{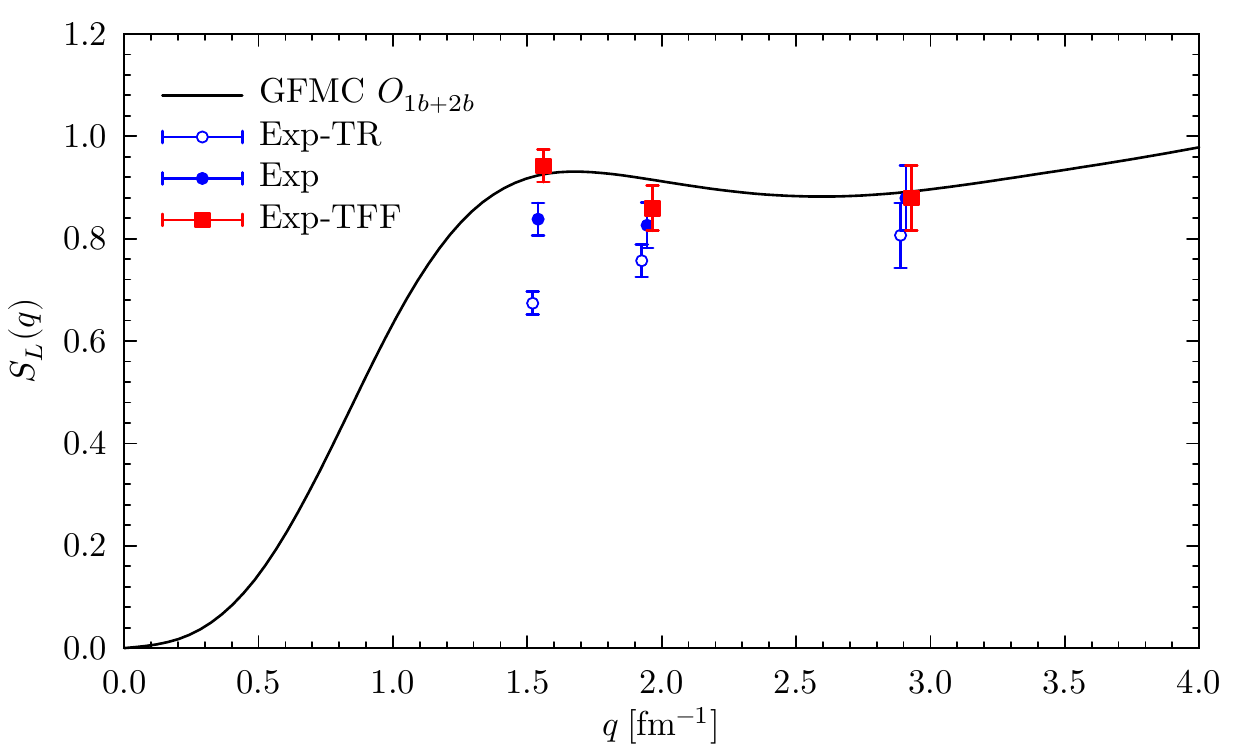}
\caption{(Color online) Coulomb sum rule in $^{12}$C: theory (black solid line labeled 1b+2b)
and analyses of experimental data (blue empty and full circles labeled EXP-TR and EXP) are from
Ref.~\cite{Lovato:2013}; the (red square) data points, labeled EXP-TFF, include the contributions
of the low-lying $J^\pi\,$=$\,2^+$, $0^+_2$ (Hoyle), and $4^+$ states, see text for explanations.}
\label{fig:f4}
\end{figure}
\end{center}

We conclude by updating in Fig.~\ref{fig:f4} the results for the
Coulomb sum rule of $^{12}$C obtained in Ref.~\cite{Lovato:2013}.
The theoretical calculation (solid line) and analyses of the experimental
data (empty and full circles) are from that work.  We recall that
the empty circles are obtained by integrating $R_L(q,\omega)$ up
to $\omega_{\rm max}$, the highest measured energy transfer,
while the full circles also include the ``tail'' contribution for $\omega > \omega_{\rm max}$
and into the time-like region ($\omega > q$), which cannot be accessed
in $(e,e^\prime)$ scattering experiments, by assuming that the longitudinal
response in $^{12}$C is proportional to that of the deuteron~\cite{Lovato:2013}.
As the direct calculations demonstrate in Figs.~\ref{fig:f1}--\ref{fig:f3},
there is non-vanishing strength in the time like-region (see in particular the
top panels of these figures which extend to $\omega > q$), and this strength needs to be accounted for before
comparing theory to experiment.

The square data points in Fig.~\ref{fig:f4} have been obtained by adding to the
full circles the contribution due to the low-lying $J^\pi\,$=$\,2^+$, $0^+_2$,
and $4^+$ states.  Given the choice of normalization for $S_L(q)$ in
Fig.~\ref{fig:f4}, this contribution is simply given by the sum of the
squares---each multiplied by $Z\,$=$\,6$---of the (longitudinal) transition form
factors listed in Table~\ref{tab:ff}.  Among these, the dominant is
the form factor to the $2^+$ state at 4.44 MeV excitation energy. 
The contributions associated with these states, in particular the $2^+$,
were overlooked in the analysis of Ref.~\cite{Lovato:2013} and, to
the best of our knowledge, in all preceding analyses---the difference
between total inelastic and quasi-elastic strength alluded to earlier was not fully appreciated.
While they are negligible at large $q$ (certainly at $q\,$=$\,570$ MeV/c), they are
significant at low $q$.  They help to bring theory into excellent agreement
with experiment.

Figures \ref{fig:f1} and \ref{fig:f3} clearly demonstrate that the picture of
interacting nucleons and currents quantitatively describes the electromagnetic
response of $^{12}$C in the quasi-elastic regime.  The key features necessary
for this successful description are a complete and consistent treatment of initial-state
correlations and final-state interactions and a realistic treatment of two-nucleon
currents, all fully and exactly accounted for in the GFMC calculations.  In the
transverse channel the interference between one- and two-body current (schematically,
1b-2b) contributions is largely responsible for enhancement in the quasi-elastic peak, while
this interference plays a minor role at large $\omega$, where 2b-2b contributions become
dominant.  The absence of explicit pion production mechanisms in this channel
restricts the applicability of the present theory to the quasi-elastic region of $R_T(q,\omega)$,
for $\omega$'s
below the $\Delta$-resonance peak.  Finally, the so-called quenching of the longitudinal response
near the quasi-elastic peak emerges in this study as a result of initial-state correlations
and final-state interactions.

\acknowledgments
A critical reading of the manuscript by Ingo Sick is gratefully
acknowledged.
This research is supported by the U.S.~Department of Energy, Office
of Science, Office of Nuclear Physics, under contracts DE-AC02-06CH11357
(A.L.~and S.C.P.), DE-AC52-06NA25396 (S.G.~and J.C.), DE-AC05-06OR23177
(R.S.), and by the NUCLEI SciDAC and LANL LDRD programs.
Under an award of computer time provided by the INCITE program,
this research used resources of the Argonne Leadership Computing
Facility at Argonne National Laboratory, which is supported by the
Office of Science of the U.S.~Department of Energy under contract
DE-AC02-06CH11357.  It also used resources 
provided by Los Alamos Open Supercomputing, by Argonne LCRC,
and by the National Energy Research Scientific Computing Center,
which is supported by the Office of Science of the U.S.~Department
of Energy under contract DE-AC02-05CH11231.

\bibliography{biblio}

\end{document}